\documentstyle[11pt,moriond,epsfig]{article}

\begin{document}
\vspace*{4cm}
\title{TODAY'S COSMOLOGICAL CONSTANTS}

\author{ R.G. CARLBERG}

\address{Carnegie Observatories, 813 Santa Barbara Street,
Pasadena, CA 91101, USA \\
{\it on leave from:}\\
Department of Astronomy, University of Toronto,
Toronto, M5S 3H8, Canada}

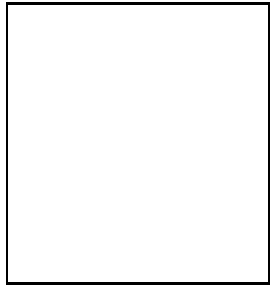
\maketitle

\abstracts{ Knowledge of the constants that describe the current
cosmological world model, $H_0$, $t_0$, and the three $\Omega$s is
central to physical cosmology.  Although there is a vast range of
suggested tests and existing constraints much of the recent discussion
of cosmological constants involves three time variable photospheres:
Cepheids, SNIa, and the CMB last scattering surface. These concluding
remarks for the Moriond XXXIII meeting are made at a time when many of
the established methods have made careful, interesting, statements
about the values of various cosmological constants based on data of
small random errors. The flood of new data over the next few years
will lead to a satisfying increase in the precision of both direct and
model dependent estimates of the main cosmological parameters. }

\section{Introduction}

Commenting on the progress being made in estimating the values of the
cosmological parameters has perils not unlike those of critiquing a
great artwork, perhaps an opera, while it is being staged for the
first time. The problem is that there is no working consensus for the
values of the cosmological constants. There are rather wide compromise
ranges which accommodate most of the derived values. However the
values of the cosmological constants at one end of a compromise range
are completely incompatible with those at the other, both in physical
meaning and in stated measurement error.

An aspect of cosmological parameter determination that is fascinating
for any interested scientist is that it continues to reward an
integrated view of the entire subject from trigonometric parallaxes to
photometric response functions to radio interferometric mapping to
X-ray plasma analysis to the outer limits of particle theory, to name
but a few. Martin Rees has described Cosmology as the ``Grandest of
the Environmental Sciences'' \cite{rees} which serves to remind us of
the difficulty of relating the observations of the universe to the
simple cosmological models of interest.  Although the FRW model and an
interest in its parameters have been around for a long time, it was
not until the 1980's when development of solid state detectors of very
low noise and very high quantum efficiency allowed virtually every
waveband used by astronomy to greatly increase both the quality and
the abundance of data. The objects now examined range from nearby
white dwarfs to galaxies at redshifts beyond 5, to precise
measurements of the CMB radiation all over the sky.  

The following discussion discusses the broad comprise ranges for the
basic cosmological parameters. In most cases the data have small
random errors.  The problems for all methods come in calibration and
model uncertainties.  For instance Cepheids distances have random
errors of about 1\%, nevertheless the total error in the Hubble
constant is generally quoted to be at least 10 times larger as a
result of potential calibration and systematic errors. Even those
apparently large error budgets are hard won from vast efforts to
control, measure and remove systematic errors.  The true error ranges
of these are continuing to shrink significantly and steadily, such
that sometime in the next decade our understanding will undergo a
phase transition, hopefully a crystallization not a meltdown.

\section{$t_0$}

The age of the Universe is currently most accurately measured from the
Globular Clusters in our own galaxy. This is an impressively mature
subject which interprets very high precision photometry using
immensely detailed physical models of relatively simple stars.  There
have been two slightly startling developments in the past few years.
First, when allowance was made for helium diffusion \cite{pm} the ages
of globular clusters came down about 10\%.  Second, the Hipparcos
satellite provided new distances to local subdwarf stars which can be
used to calibrate the distance to similar stars in globular
clusters. The surprise was that the distances were somewhat larger
than anticipated, meaning that the stars were brighter and hence
younger.  The quoted ages are $11.5\pm1.5$~Gyr \cite{cdkk} and a
minimum of $13$~Gyr \cite{pmtv}.  The physical complexities of
confidently predicting the model luminosities in the observed filter
bands leads to some further uncertainty. Although the results are
statistically compatible, within this range one can draw quite
different cosmological conclusions.  Since one generally adds 1~Gyr to
the Globular cluster ages to derive a $t_0$ it appears that the
traditional first year astronomy values of 12 to 15~Gyr remain
acceptable values. There is great excitement that white dwarfs will
soon provide an alternate dating method to some of the nearby globular
clusters.

\section{$H_0$}

In principle, measuring $H_0$ is simple. In practice it is a
dimensional constant which effectively requires carrying the standard
meter from the earth to cosmological distances where the fractional
perturbations to the Hubble flow are small.  The classical methods are
entirely geometric, relying on parallax and the $1/d^2$ law, but
require building a ``distance ladder''.  A variety of newer methods
are single step, but, model dependent. For instance, combining cluster
X-ray and Sunyaev-Zeldovich measurements, or, gravitational lenses
with measurable time delays between images, gives a quantity that
depends on the Hubble constant.

Classical estimation of $H_0$ has been advanced enormously with CCD
photometry and telescopes with greatly improved angular resolution,
allowing the brightness of point-like objects to be accurately
measured at relatively large distances.  Cosmology has always been a
matter of dealing with state of the art detection of photons from
distant, faint objects which are themselves the focus of research.  It
is not surprising that it has been slow progress since Hubble's
original discovery (whose 1936 book gave a well considered Hubble
constant of about 600 km~s$^{-1}$~Mpc$^{-1}$). The Hubble constant
search has become a substantial industry with many approaches to the
problem. The HST Key Project \cite{keyproject} to measure Cepheid
distances to a representative set of objects in the local supercluster
is generally regarded as a piece of bedrock which is a standard rod
against which other measurements approaches are calibrated or
compared. Although the distances to the various galaxies that have
been measured are fairly non-controversial, the implied Hubble
constant remains quite controversial. It would be desirable if more
SNIa would co-operate and explode in some of these galaxies with
carefully measured disteances.

It is impressive that in about the last 10 years the scatter in the
values of the Hubble constant have decreased by a factor of two or
perhaps a little more\,\cite{kfm}. Many would accept that the Hubble
constant lies someplace in the range of 55 to 75
km~s$^{-1}$~Mpc$^{-1}$.  No one is really comfortable with such a
large range.  The good news is that the Key Project has more results
to announce, and the rate at which data are being collected in other
approaches to this problem is increasing.  The next 10 years will see
total errors reduced another factor of two, which should lead to
consensus of some sort.

Newer methods like the SZ/X-ray technique \cite{sz,myers} or the
gravitational lens time delay measurements \cite{ws} are promising
one-step methods with simple physics, however in both cases the
results at the moment have a fairly large scatter compared to
classical estimators and there remain genuine complications in
modelling the potentials involved in the cluster potentials and the
the lensing potentials. On the other hand, the interest, effort and
data quality of these new methods are ascending very rapidly.

\section{The Three $\Omega$s}

The three controversial $\Omega$s are the $\Omega_B$, $\Omega_X$, and
$\Omega_\Lambda$, the density parameters of the baryons, dark matter
and the $\Lambda$ term, respectively. The matter density,
$\Omega_M=\Omega_B+\Omega_X$, is the total density of gravitating
matter which has no significant pressure. The fourth $\Omega$ is that
of radiation, which although by far the smallest at the current epoch,
is known to fabulously high precision through the measurement of the
spectrum of the CMB radiation.

Over the past 10 years there has been a great deal of new work on the
baryonic $\Omega_B$ which probably hasn't changed the formal error
range that much, but, many systematic errors have been reconsidered.
The range of quoted values is about a factor of three from the bottom
end of the ``Helium'' value to ``low Deuterium'' value,
$\Omega_B=0.007-0.024h^{-2}$ \cite{st}. Helium is particularly
difficult because one needs to derive the primordial abundance from
systems that have been partially contaminated with stellar
Helium. Deuterium is both very fragile in stars and difficult to
observe. Galaxy clusters estimate the baryon fraction, the ratio of
the gas plasma baryonic mass to total mass within some radius,
$f_B\simeq 0.12\pm0.02h_{50}^{-3/2}$ \cite{wjf}, which is consistent
with the BBN $\Omega_B$ provided that $\Omega_M$ is less than
approximately 0.4. This suggests, but does not yet prove, that
clusters contain a representative gas-to-mass ratio.  

An extremely significant, and confident result, is that although
$\Omega_B$ is about an order of magnitude larger in gas than in stars,
it is still about an order of magnitude less than the total
gravitating mass. Hence, the unknown dark matter continues to be the
dominant mass in the universe.  Another puzzle is that the current
location of most of the baryons is not known, with the IGM being the
usual suspect, although MACHOs are a new possibility.

The total $\Omega_M$ has seen a lot of variation over the past 20
years, coming back to rest someplace near the views of the classic
article of the mid-1970's \cite{ggst}, which argued for an
$\Omega_M\simeq 0.1$. In spite of the apparent lack of change this
subject has gone the full circle with tremendous profit to
astrophysics.  The circle began when Peebles derived the Cosmic Virial
Theorem \cite{cvt} which helped to provide the motivation for large
redshift surveys. One of the first redshift surveys was the CfA, which
indicated $\Omega\simeq 0.2$ \cite{dp}, under the assumption that
galaxies trace the mass. About the same time there were two
theoretical developments which lead to hesitation in accepting this
result. First, inflation \cite{guth} argues very persuasively that
$\Omega_M=1$. Second, the concept of formation of objects at peaks of
the density fluctuation field immediately demonstrated that galaxies
could be clustered much more strongly than the underlying total mass
\cite{kaiser}, hence the derived $\Omega_M$ could be severe
underestimates.  Virtually all theorists concluded that in the absence
of overwhelming evidence to the contrary, $\Omega_M=1$.  A little
later there was the largely unanticipated discovery of large scale
flows of galaxies \cite{7s} which appeared to be a very strong
argument for $\Omega_M\simeq1$.

In the past few years, the $\Omega_M$ tide has reversed to generally
favouring lower $\Omega_M$ values again.  That is, evidence for a low
$\Omega_M$ that strikes many as overwhelming is gradually being
accumulated.  A prediction of high $\Omega_M$ is that clusters should
have an increasing M/L with radius, as the Milky Way galaxy does
beyond the disk. However, X-ray, redshift survey and weak lensing data
all find that cluster galaxies are distributed like the dark matter to
the virial radius.  Furthermore, as newer catalogues of peculiar
velocities and nearly-all sky surveys become available it appears that
the flow fields are actually consistent with $\Omega_M$ as low as 0.2
or so \cite{davis,dacosta}.  Galaxies cannot be completely unbiased
with respect to the total matter content of the universe, since
various galaxy types and luminosities have different clustering
properties. However, it is gradually coming clear that the bias cannot
be very large.  Much of the community would agree that $\Omega_M$ is
someplace in the range $0.1-0.4$.  Direct dynamical data favour
$\Omega\simeq0.2$, but CMB fitting suggests the higher values. Now
that much of the community has adopted a low density universe as the
most likely model, there is a growing tension between very low
$\Omega_M$, those less than 0.2, and moderate $\Omega_M$, those
greater than 0.3, which point to some unresolved systematic error in
the various data involved.

The most uncertain, and puzzling, of the $\Omega$s is
$\Omega_\Lambda$.  Here the measurement situation is far from
mature. An upper limit on $\Omega_\Lambda$ comes from the number of
split images of distant sources produced by gravitational lensing from
intervening galaxies. The very few splittings seen argues that
$\Omega_\Lambda$ is less than about 0.7.  This argument hasn't changed
a lot since the 1980's \cite{turner} but it has become a lot more
secure as worries about optical selection and evolution in the lensing
galaxies have been addressed \cite{kochanek}. The preliminary evidence
from the CMB fluctuation measurements (see various contributions to
this meeting) is that if $\Omega_M$ is low there must be a nonzero
$\Omega_\Lambda$ to produce a peak in the angular power spectrum at a
sufficiently large angle. 

A scientific banquet of ground based, balloon borne and satellite CMB
experiments is beginning, such that this meeting stands at a moment
when exponential growth of effort is happening in the kitchen, with
the expectation that a feast of high quality results begin in about a
year. Many of these experiments have made impressive claims for the
precision of their measurements, which are certainly true if the
instruments perform to specification and the universe and the
foregrounds correspond to the various model assumptions.

A completely classical approach to $\Omega_\Lambda$ is the
magnitude-redshift relation, which is underway with the SNIa
measurements. These objects can be detected and reliably studied up to
redshift of one, which provides considerable leverage on the
$\Omega_M-\Omega_\Lambda$ pair. The obvious concern is that some
unknown evolution or systematic measurement error could creep in. The
two teams are well aware of these possibilities and appear to be
recalibrating, worrying about the astrophysics and the astronomy as
much as anyone else. As much as joint analyses have the power to
produce better parameter estimates, one looks forward to independent
methods giving completely independent results with small errors, that
do agree.

It is impossible to avoid a philosophical note when contemplating the
$\Omega$ values. That is surely the point. On the other hand, the
universe can do what it wants and it is our job to measure and try to
understand the results, which is a lesson from the history of particle
physics.  However, one cannot help but note the conundrums of a
$\Lambda$ dominated universe. If $\Omega_\Lambda$ was much larger than
unity, the universe would have started to exponentiate before galaxies
(and presumably most stars) ever formed. Similarly even if
$\Omega_\Lambda+\Omega_M=1$, then we are currently well into a round
of inflation. We might ask about the prospects for our scientific
descendents. Taking a rather modest extrapolation, by astrophysical
standards, of 20 times the current age of the universe, then at that
future time there will have been about 20 e-folds of expansion, so the
CMB temperature will be nearly zero, the local group will merge with a
few satellites left orbiting, and, all other galaxies will have
receded to unobservable redshifts. The result is that future
astronomers are left in a universe that has one significant galaxy:
the merger product of ours and M31.  Most peculiar, but, evidently
possible within this model framework.

\section{Conclusion}

The current situation is precisely why the measurement of the
cosmological parameters is the primary activity of many astronomers
and astrophysicists. The exciting likelihood is that most of the major
cosmological constants will be known to a satisfying degree of on the
time scale of a decade.

The Moriond meetings provide an ideal format for frank discussion of
extremely controversial cosmological issues. I thank the organizers
for the splendid job they did in managing to bring us all together and
providing a never ending flow of food and stimulus for mind and body.

\end{document}